# A Hedonic Metric Approach to Estimating the Demand for Differentiated Products: An Application to Retail Milk Demand

Osman Gulseven[1] and Michael Wohlgenant

This article introduces the Hedonic Metric (HM) approach as an original method to model the demand for differentiated products. Using this approach, initially we create an n-dimensional hedonic space based on the characteristic information available to consumers. Next, we allocate products into this space and estimate the elasticities using distances. What distinguishes our model from traditional demand models such as Almost Ideal Demand System (AIDS) and Rotterdam Model is the way we link elasticities with product characteristics. Moreover, our model significantly reduces the number of parameters to be estimated, thereby making it possible to estimate large number of differentiated products in a single demand system. We applied our model to estimate the retail demand for fluid milk products. We also compared our results with the Distance Metric (DM) approach of Rojas and Peterson (2008) using the estimation results from traditional models as a benchmark point. Our approach is shown to give superior results and better approximations to original models.

---

[1] Osman Gulseven is an Assistant Professor, Department of Economics, Middle East Technical University, Ankara, 06531, TURKEY, Email:gulseven@metu.edu.tr, Phone: 90-312-2103043, Fax: +90-312-217964; Michael Wohlgenant is William Neal Reynolds Distinguished Professor, Department of Agricultural & Resource Economics, North Carolina State University, Raleigh, North Carolina 27695-8109

# A Hedonic Metric Approach to Estimating the Demand for Differentiated Products: An Application to Retail Milk Demand

A significant amount of empirical research concentrates on estimating price elasticities of demand between products of similar types. In traditional models such as the Almost Ideal Demand System (AIDS) and the Rotterdam model, the time-series relationship between prices and market shares is exploited to estimate the own-price, cross-price and income elasticities. Although theoretical restrictions reduce the number of parameters to be estimated, as the number of goods gets large, the number of parameters to be estimated increases exponentially.

In recent years, random utility models have been a popular alternative to estimating demand elasticities. In this approach, the utility maximizing consumer chooses the products that give the maximum utility derived from the commodity attributes. Random utility models that extend the simulated maximum likelihood approach of Berry, Levinsohn, and Pakes (1995) reflect characteristic differences in elasticities yet they are very computationally complex to estimate. (Hendel 1999; Nevo 2001; Chan 2006) A simpler Distance Metric (DM) approach by Pinske, Slade, and Brett (PSB, 2002) uses spatial distances to estimate price elasticities between different locations. Their model is simple enough to be estimated without simulations, yet flexible enough to characterize the substitution patterns between differentiated products.

Rojas and Peterson (2008) apply this new method to the retail beer market using alcohol content as the main distance measure along with different distance combinations. However, their choice of distances is ambiguous and depends on prior judgments about the data. The Hedonic Metric (HM) method proposed in this article alleviates this ambiguity while reducing the number of parameters to estimate.

Our methodology is based on a two-stage estimation technique. In the first stage, we estimate a hedonic equation to obtain the price of each attribute available in the product. For retail markets, the information on the labels is a good choice for attributes. The differences in the attributes of product varieties are exploited to create a hedonic matrix based on hedonic distance. These measures are based on pair-wise comparisons of Euclidean distances where the amount of each characteristic in the product is weighted by its hedonic price. Once we allocate each differentiated product into the hedonic space, the distances are used to estimate the demand elasticities.

For comparison, the model was also estimated using the DM approach of Rojas and Patterson (2008). The models are evaluated through comparison of the elasticities estimated using the retail milk market data. Each metric method (HM, DM) is evaluated with regard to closeness of its elasticities to the original model where elasticities are not approximated by either method. Our method is shown to outperform the DM methodology. Moreover, the results were found to be close to the original model whether we used the AIDS or Rotterdam model. Given space restrictions, we report only the results for the Rotterdam model.

**Model Specification**

The Rotterdam model is derived by totally differentiating the Marshallian demand functions and substituting the Slutsky equation to derive the relationship between market shares and prices in a demand system such that for *n* goods:

$$w_i d \log q_i = b_i d \log X + \sum_j c_{ij} d \log p_j \quad \text{where } i = 1\ldots n \text{ and } j = 1\ldots n \qquad (1)$$

$w_t = 0.5(w_t + w_{t-1})$, $d \log X_t = d \log X_t - \sum_j w_t d \log p_{jt}$, $d \log p_{jt} = \log p_{jt} - \log p_{jt-1}$ and $X_t$ refers to the total expenditure on products at time $t$. Theoretical restrictions can be imposed in the form of adding up, $\sum_j b_i = 1 \ \forall i$, $\sum_i c_{ij} = 0 \ \forall j$; homogeneity, $\sum_j c_{ij} = 0 \ \forall i$ ; and symmetry, $c_{ij} = c_{ji} \ \forall i, j (i \neq j)$. Dividing each $c_{ij}$ by the average own market share gives the compensated (Hicksian) price elasticity $(e_{ij} = c_{ij}/w_i)$; dividing each $b_i$ by its average own market share gives the expenditure elasticity $(e_i = b_i/w_i)$; and uncompensated (Marshallian) demand elasticities are recovered using the Slutsky equation $\left(e_{ij}^m = e_{ij} - e_i w_j\right)$.

*Distance Metric Method*

The distant metric approach used by Rojas and Patterson is an approximation method to estimate the relationships between prices and market shares. The cross price coefficients $(c_{ij} \; \forall \; i,j)$ are defined as functions of distant measures between products such that

$$c_{ij} = \sum_{l=1}^{L} \lambda_l d_{ij}^l \qquad (2)$$

where $L$ is the number of attribute spaces and $d_{ij}^l$ is the distance between product $i$ and $j$ in space $l$. These distant measures are based on the product attributes and can be continuous (content), discrete (type) or both. The continuous distance measures are defined as the inverse measure of Euclidean distance in attribute space between products. In this form, these measures range between 0 and 1. Since the inverse of the distance measures are used for parameterization, the distance measure refers to the closeness of these products such that a higher index (close to 1) implies closer products whereas a lower measure (close to 0) implies distant products. For an n-dimensional attribute space, the distance $l$ between two products $i, j$ can be defined as

$$d_{ij}^l = \frac{1}{1 + \sqrt{\left(\delta_{ij}^1\right)^2 + \left(\delta_{ij}^2\right)^2 + .. + \left(\delta_{ij}^n\right)^2}} \qquad (3)$$

where $\delta_{ij}^k$ is the distance measure in dimension k. For example, the closeness index between 2% milk and 1% milk based on a three-dimensional fat-organic-size (FOS) attribute space is calculated as

$$d_{2\%,1\%}^{FOS} = \frac{1}{1 + \sqrt{\left(\delta_{2\%,1\%}^{FAT}\right)^2 + \left(\delta_{2\%,1\%}^{ORGANIC}\right)^2 + \left(\delta_{2\%,1\%}^{SIZE}\right)^2}} \qquad (4)$$

The continuous distances between products are scaled by dividing the differences in contents with the maximum amount of content available in any product. For example, the fat distance between 2% Milk and 1% Milk is calculated as

$$\delta^{FAT}_{2\%,1\%} = \frac{Fat\ Content\ of\ 2\%\ Milk - Fat\ Content\ of\ 1\%\ Milk}{Fat\ Content\ of\ 3.25\%\ Milk} \quad (5)$$

The discrete distances can be based on type of the product depending on whether they are in the same classification or not. In that form, they are either equal to 1 (same classification) or 0 (different classification). Also, a product is the nearest neighbor (NN) of another product if it has the highest closeness index for a given attribute space. By construction, the discrete distance measures are normalized to one such that for a given attribute space, the sum of distance measures for each product type equals one.

The own-price coefficients are also specified in terms of product attributes. However, this time, the actual product attributes interact with the own-price coefficients.

$$c_{ii} = \beta_0 + \sum_{j=1}^{C} \beta_j \chi_i^j \quad (6)$$

where $\beta_0$ is the constant coefficient on the own-prices and $\chi_i^j$ is the content of characteristic $j$ in product $i$ that interacts with the price.

Incorporating these parameter approximations to the Rotterdam model gives us the following empirical Distance Metric approximated Rotterdam model (DM-RM):

$$w_i d \log q_i = b_i d\log X + \beta_0 \operatorname{dlog} p_i \\ + \sum_{k=1}^{C} \beta_k \chi_i^j \operatorname{dlog} p_i + \sum_{j \neq i} \sum_{l=1}^{L} \lambda_l d_{ij}^l\ d \log p_j \quad \forall i,j \quad (7)$$

Since different approximation techniques are used for own-price and cross-price coefficients, the corresponding elasticities are calculated based on the approximation technique: The Hicksian own-price elasticity, $e_{ii} = \beta_0 + \sum_{j=1}^{C} \beta_j \chi_i^j / w_i$ ; the Hicksian cross-price elasticities, $e_{ij} = \sum_{l=1}^{L} \lambda_l d_{ij}^l / w_i$; Income (expenditure) elasticities are calculated as in the original Rotterdam Model as $e_i = b_i / w_i$ ; and the Marshallian elasticities are recovered using the Slutsky equation in elasticity form.

*Hedonic Metric Method*

The hedonic metrics proposed in this article is a better alternative to the distance metric estimation since we alleviate the ambiguity in the choice of distances. In both approaches, the elasticities between differentiated products are approximated using the distance measures between products. However, in the DM approach these distances are based on pre-imposed specific attributes, whereas we allocate each product in the multidimensional hedonic space. In order to create distances between products, first a hedonic regression is estimated to get the hedonic prices of each attribute.

According to hedonic theory, each consumer is trying to maximize his/her own utility that depends on the product attributes (Lancaster 1966). Therefore, the consumer maximizes utility by selecting products that maximize the sum of utilities derived from each attribute (Rosen 1974). Based on the hedonic model, the price of each good can be characterized by the set of its attributes that comes with the product. Defining this set as $x = [x_1, \ldots, x_k]$, the functional relationship between the price of a good and its characteristics vector $X$ can be stated as $p = f(x) + \mu$ where $\mu$ is the error vector.

If the relationship between prices and attributes is assumed to be linear then the price of a good $i$ can be derived as the sum of the attribute values (Ladd and Suvannunt 1976). Thus the retail price of the product is equal to the sum of monetary values of product attributes where the total value of each attribute is equal to the quantity of the attribute multiplied by the implicit price of that attribute. This implies

$$P_i = \sum_{j \in J} x_{ji} \beta_j + E_i + \epsilon_i \quad J \text{ is the set of product attributes} \tag{8}$$

where $x_{ji}$ is the amount of attribute $j$ in product $i$ and $E_i$ is the unique characteristic of the product. The implicit prices of characteristics can be calculated as the partial derivatives of the hedonic functions.

$$\frac{\partial p}{\partial x_j} = \frac{\partial f(x)}{\partial x_j} = \beta_j \quad \forall j \tag{9}$$

Thus, in the linear model, the coefficients on the attributes give us the hedonic prices for these attributes. The value added for each attribute is calculated by multiplying the implicit price by the attribute quantity. For product $i$, the value added from attribute $j$ is $v_{ij} = x_{ji} \beta_j$.

If the price attribute relationship is assumed to be in semi-log form (Nimon and Beghin 1999), then instead of price, the log-price of the product is defined in terms of attributes such that

$$\log P_i = \sum_{j \in J} x_{ij} \beta_j + E_i + \epsilon_i \tag{10}$$

In this form, the implicit price of the attribute is calculated by multiplying the coefficients on attributes with the price of the products

$$\frac{\partial p}{\partial x_j} = \frac{\partial f(x)}{\partial x_j} P_i = \beta_j P_i \ \forall \, j, i \tag{11}$$

The value added term also accounts for the price of the product $v_{ij} = x_{ji} \beta_j P_i$. The semi-log form implies that the same amount of attribute can have a higher value if it is located in a product with a higher retail price.

The difference in the value added for each product is used to calculate hedonic distance in terms of a single attribute. Combining the sum of these price-weighted attribute distances and rescaling them to be between 0 and 1 gives us the continuous hedonic distance matrix. Similar to distance metrics, two products are nearest neighbors (NN) if they have the highest closeness index in hedonic space. Thus, the nearest neighbor concept is introduced as a discrete distance based on the hedonic distance matrix. The nearest neighbor matrix along with the hedonic distance matrix define the cross-price coefficients such that

$$c_{ij} = \lambda_h d_{ij}^h + \lambda_{nn} d_{ij}^{nn} \tag{12}$$

where $d_{ij}^{\,h}$ refers to the hedonic distance and $d_{ij}^{\,nn}$ refers to the distance based on the nearest neighbor concept. Own-price coefficients are also approximated by interacting product attributes with own-prices. They are estimated based on each product's average market share and inverse of the hedonic distance vector (i.e., closeness index)

$$c_{ii} = \beta_0 + \beta_1 \chi_i^s + \beta_2 \chi_i^c \tag{13}$$

where $\chi_i^s$ refers to the market share and $\chi_i^c$ is the closeness index of product $i$.

Putting these coefficients back in the original model, we get the following Hedonic Metric approximation to the Rotterdam model (HM-RM):

$$w_i d\log q_i = b_i d\log X + \beta_0 \, d\log p_i + (\beta_1 \chi_i^s + \beta_2 \chi_i^c)\, d\log p_i + \sum_{j \neq i} (\lambda_h d_{ij}^h + \lambda_{nn} d_{ij}^{nn})\, d\log p_j \quad \forall i,j \tag{14}$$

In this form, the Hicksian own-price elasticity is $e_{ii} = (\beta_0 + \beta_1 \chi_i^s + \beta_2 \chi_i^c)/w_i$; the Hicksian cross-price elasticities can be calculated as $e_{ij} = (\lambda_h d_{ij}^h + \lambda_{nn} d_{ij}^{nn})/w_i$; Income elasticities are calculated as in the original model as $e_i = b_i/w_i$; and the Marshallian elasticities are recovered using the Slutsky equation in elasticity form.

**Data**

Nutritional food attributes derived from the Nutrient Database and the marketing attributes derived from Homescan Data characterize milk products in hedonic space.

Most of the attributes in Homescan data are based on nutritional claims such as organic label, soymilk, lactose free, cholesterol free, vitamin enrichment, and calcium enrichment. However as Stranieri, Baldi, and Banterle (2010) suggest, a significant ratio of consumers use nutritional labels while making purchase decision. Therefore information about attributes is enriched using the USDA Nutrient Database. In particular, continuous nutritional contents including protein, carbohydrate, fat amount, along with sodium and cholesterol contents are obtained. Combining these two databases gives us an exact set of attributes that are also available on the product labels.

While major nutritional components are measured in grams per serving, measurement units for vitamins and mineral are defined in different units. We transformed all values for vitamins and minerals into daily recommended intake (DRI) percentages provided in a serving size defined by the U.S. Federal Drug Administration (2008). The vitamin and mineral contents other than sodium are combined into a single vitamin-mineral index. Sodium and cholesterol are two important components listed on the product labels that may raise health concerns among consumers (Garretson and Burton 2000; Chema et al. 2006; Moon, Balasubramanian and Arbindra 2005). In fact, studies show that they have a negative effect on the value of the product and consumers' preferences (Harbor-Locure, McLean-Meyinsse and V. Bethea 2001; Peng, West and Wang 2006). Therefore, we excluded sodium from this index and included it as a separate component due to its differential effect.

*Product Attributes*

The final attribute space includes fat type, organic claim, soy dummy, promotion dummy, lactose/cholesterol free (LFCF), vitamin-mineral enhancement, and nutritional variables such as protein, carbohydrate, lipid (fat) content, percentage daily recommended intake (DRI) index of cholesterol, sodium, and vitamin/minerals along with the servings per package.

The summary values for all components are given in table 1**.** Each milk category other than soy milk includes hundreds of different Universal Product Codes (UPC) where each UPC differs from another by at least one characteristic. While the dominant characteristic difference between different milk types is fat content, we observe a large variation in product characteristics between milk types and also within each milk type as well. Soymilk has a much higher organic ratio compared to other types of milk. It is also promoted more than other milk types and it is lactose/cholesterol free. Another distinguishing difference is in average serving size per package. On average, soymilk comes in smaller packages than cow's milk. Among the dairy based milk types, skim milk includes the most differentiated milk purchases. It has the highest protein per serving, and also highest ratio of vitamin-mineral enhancement. Moreover, almost 4% of skim milk is lactose-cholesterol free. Since lactose is naturally present in dairy products, the summary results show that skim milk is the most functionally enhanced type.

*Aggregate Data*

While it is preferable to use individual level purchase data to analyze household purchase behavior, we used aggregate data. It is possible to estimate the household level demand in traditional models and also hedonic metric approximations, but it is impossible to use distance metrics at the household level. The distance metric space is based on product characteristics which are different for each consumer. It will be extremely cumbersome to find the most feasible distance metric measures at the household level since each household would allocate the products in a different metric space. Moreover, whichever attribute list we use, there will be many attributes missing in distance space which creates another problem in estimation. Since our primary aim is to introduce hedonic metrics and compare them with distance metrics, we aggregated our data into weekly purchases made by all core households.

The data contain information on weekly prices, quantities and expenditures on five different milk types between 2002 and 2005. The average market share of 2% Milk is highest with a 34% share, followed by Skim Milk with a 27% market share. Whole Milk (3.25%) and 1% Milk each have 18% market shares on average. The price of soymilk per serving is almost double that of other milk types. Because soymilk is primarily organic, lactose free, cholesterol free, and consumed in smaller (more convenient) packages, it has more desirable attributes than other milk types. Table 2 gives the price-quantity-market share statistics for each milk type.

*Distances*

The data on prices and market shares are enough to estimate the price and income elasticities in the context of traditional demand models. However, defining distances is the most crucial part of DM and HM approximations. In this article, not only the cross-price elasticities are recovered from these distance coefficient estimates, but so are the own-price elasticities. The most distinguishing characteristics of milk types are organic percentage, fat content, and container size. Thus, distances based on these characteristics are used in DM estimation.

Purchasing organic milk could be a discrete decision for an individual for a specific purchase occasion. However, with aggregate data, organic percentage is another dimension that captures the substitution effects between milk products in organic market. If two milk types have both higher organic percentage ratios, they will be closer substitutes in the organic milk market and our model implements this interaction. The same idea applies to the size variable. Purchasing a specific size could also be a binary decision for a consumer at a specific occasion. However, in aggregate form, the size variable is an important dimension that shows the degree of substitution in different package sizes (Kumar and Divakar 1999).

The continuous distance measures can be single dimensional based on these attributes (Fat, Organic, Container), two-dimensional based on pair-wise distances (Fat-Organic, Fat-Container, Organic-Container), or three dimensional (Fat-Organic-Container). The distance measures are calculated from the differences between milk types based on these measures. Following PSB's methodology, a discrete distance based on nearest neighbor (NN) concept is also introduced for two-dimensional and three-dimensional distances. Two products are nearest neighbors if they are next to each other in the attribute space. For the purpose of approximating own-price elasticities in DM, the contents based on market share (W), fat content (F), organic claim (O) and servings per package (S) are used to approximate these elasticities.

The hedonic metric approach proposed in this article also uses distances between product attributes to approximate the cross-price and own-price elasticities. The distinction between these approaches is the distances used in the estimation. In the HM approach, the products are located in the hedonic space which characterizes them. To get the location of each product type, implicit prices of the attributes that characterize the product are estimated using hedonic regressions. Using these implicit prices, the value added of each attribute is calculated for each milk type. For a linear hedonic model, the estimated coefficients in attribute quantities give us hedonic prices. For a semi-log model, these coefficients are multiplied by the prices to obtain the hedonic prices.

The difference in value added scaled by the prices of the attributes gives us the location of each product in hedonic space. Thus, the hedonic prices of the attributes scale the attributes according to their values. The nearest neighbor concept in HM approach is also based on hedonic distances. For estimating the own-price elasticities, a different measure based on the sum of total pair wise distances between each product is utilized. Since the distance measure is in inverse form, this measure acts as a total closeness index based on hedonic distance. For example, because soy is the most unique product in hedonic space, it has the lowest closeness index among all milk types. This closeness index based on the inverse of hedonic distance is used along with the product's market share to estimate own-price elasticities for each milk type.

**Results**

The estimation results based on both methods closely resemble that of the original models. Because we reduced the number of parameters through approximations, the significance of elasticity estimates are higher in approximated models. However, the approximated elasticities that are derived from hedonic metrics outperform the distance metrics.

*Models without Elasticity Approximations*

The original models (without approximating elasticities with DM or HM) were estimated for the system of equations consisting of 2%, Skim, Full, 1% milk, and soymilk. The equation for soymilk was dropped prior to estimation to avoid singularity. The equations were estimated using the seemingly unrelated regressions method (SUR). The parameter estimates for the soymilk demand equation are obtained through application of the adding up, homogeneity, and symmetry restrictions. The Durbin-Watson statistics for each milk type indicate there is no autocorrelation in the residuals.

All own-price elasticities and expenditure elasticities are statistically significant and the elasticity estimates are very close to each other. As can be seen from table 3, soy milk has the highest own-price elasticity followed by skim milk, 2% milk, 1% milk and whole milk. While whole milk has the highest average price among the dairy based milk types, it has the lowest own-price elasticity.

*Distance Metric Approximations*

To identify the most distinguishing distances, first we estimated the DM approximation to the Rotterdam model using only single continuous distances one-by-one. The results from these estimations in table 4 indicate that all distances are positive as expected. Since dairy based milk types are classified according to fat content, closeness in fat content space is included in the estimation. Organic percentage is also another distinguishing attribute along with the average serving size per package that identifies the milk type in the attribute space. A variety of different nearest neighbor distance combinations have been estimated to get the closest approximations to the original models. The three different distance measures that are used in estimation are *Fat, Organic-Size, and NN for Fat-Organic-Size* space (*F-OS-NN$_{FOS}$*); *Fat, Organic-Size, and NN for Fat-Organic* space (*F-OS-NN$_{FO}$*); and *Fat, Organic, and NN for Fat-Organic* space (*F-O-NN$_{FO}$*). Since the estimation results are similar across these three versions, we report only the results for F-O-NN$_{FO}$ space based on the lowest Akaike (AIC) and Schwartz Information Criteria (BIC).

The estimated Marshallian and Expenditure elasticities for DM-RM models can be found in table 5. The estimated expenditure elasticities match closely with the traditional models where the elasticities are not approximated. All possible continuous distance combinations (*Fat, Organic, Size, Fat-Organic, Fat-Size, Organic-Size, and Fat-Organic-Size*) along with the NN distances for Fat-Organic, Fat-Size, and Fat-Organic-Size are included in the full model. The NN for Organic-Size dimension is not included since it allocates milk types in the same way they are allocated in the Fat-Size dimension. The full model gives the lowest AIC and BIC scores; however none of the estimated approximation parameters are significant. We fail to reject the equivalence of own-price elasticities for all milk types at the 5% significance level. The cross price elasticities that measure the effect of soymilk prices on the market shares of 1% milk, and full-fat milk ($e_{1\%,Soy}^m$ and $e_{Full,Soy}^m$) are statistically different from the original elasticities between these products. Moreover, the high degree of correlation between distance measures suggests use of only the most important characteristics as distances.

If we use a subset of distances based on important characteristics, it is observed that related own-price and expenditure elasticities are significantly different from zero. Moreover all approximated elasticities are within the 95% confidence interval of the original elasticities. The own-price coefficient terms based on market share, fat content, and organic claim are not statistically significant, whereas the constant price coefficients are significantly negative indicating a negative relationship between own prices and market shares.

Similar to the results for non-approximated elasticities, soy milk has the highest uncompensated own-price elasticity, yet the rest of the own-price elasticities differ significantly. The DM method underestimates the own-price elasticities for 2% milk and skim milk and it overestimates the cross-price elasticities for full milk and 1% milk. In version *F-O-NN$_{FO}$*, the order of own-price elasticities can be ranked as soy milk, followed by 1% milk, full fat milk, 2% milk and skim milk. Although the models approximate the own-price elasticities for 1% milk and soymilk well, they underestimate these elasticities for 2% milk, and skim milk and overestimate that of full fat milk. The coefficient on the organic distance is also significantly positive indicating the positive relationship between distances in organic percentage space and substitutability of the milk types regarding organic dimension.

*Hedonic Metric Approximations*

The monetary values of product attributes are estimated using both linear and semi-log hedonic regressions. The results are summarized in table 6. Regardless of the choice of model, the most distinguishing attribute is the Lactose-Free/Cholesterol-Free label followed by the Organic claim. Soy attribute is highly influential in the product price. A vitamin mineral enhancement label has a positive effect on price; however if a product is discounted, we expect to see a reduction of 1-2 cents in price per serving (or 8-16 cents per ½ Gallon package). Among the nutritional attributes, protein has the highest value followed by carbohydrate and fat content. Both cholesterol and sodium contents have significantly negative effects on the product values. Vitamin and mineral content are highly valued while an increase in size reduces the price per serving.

The estimated compensated price elasticities and expenditure elasticities along with summary of estimation results for HM-RM models are shown in table 7. Regardless of the initial hedonic regression version, the estimated expenditure elasticities match almost perfectly with the original models. All own-price elasticities and expenditure elasticities are significant and their signs are as expected. Moreover, the coefficient on the inverse of the hedonic distance terms is negative indicating products closer in hedonic space are close substitutes. The substitution effect declines as the distance between products in hedonic space increases. In addition the price interaction coefficient on the hedonic uniqueness term is also negative. This implies more unique product types have higher own-price elasticities. This is true in the case of the own-price elasticity for soy milk since soy milk has the highest uncompensated own-price elasticity among all milk types. In the linear hedonic based model, the own-price elasticity for soymilk is highest followed by that of 1% milk, full-fat milk, skim milk and 2% milk. In the semi-log hedonic based model, the own-price elasticity of full-fat milk is slightly higher than that of 1% milk yet the order of other own-price elasticities are not different than the linear model. All of the estimated elasticities (estimated at the mean values) are within the 95% confidence interval of the original Rotterdam Model results. This indicates that the Hedonic Metric based approximation to the Rotterdam model performs very well.

**Conclusions**

In this article we compare the Distance Metric (DM) method with the Hedonic Metrics method (HM) regarding their performance in approximating the elasticities estimated by the Rotterdam and LA/AIDS models. The data used in our estimation are AC Nielsen Homescan data, which record household level purchases, and the USDA Nutritional Database, which provides detailed nutritional facts about individual products. Combining these sources and aggregating consumption on a weekly basis gives us time-series quantity-price and market share data that are used to estimate demand for different milk types. The uncompensated cross-price elasticities indicate that the soymilk prices do not affect the market shares of dairy based milk types, yet the inverse might not be true.

The DM approximation based on all possible combinations of distance measures give non-conforming estimates for some cross-price elasticities which might be due to high correlation between some measures. Thus, some elasticity estimates do not fit in the 95% confidence interval of those estimated by the traditional model. Also, the order of estimated own-price elasticities based on the DM approximation is different than the original model. The significant inverse distance coefficient is positive suggesting that closer product types are closer substitutes.

The HM approximations are based on hedonic distances calculated as the sum of the pair wise differences in the value added of each attribute for each product type. Therefore it eliminates the need to search for significant characteristics and has a stronger foundation than the DM method. The calculated elasticities are very similar to the actual ones. In fact, all mean elasticities fit into the 95% confidence interval of original estimation results. The coefficient on the hedonic uniqueness parameter is negative suggesting unique products have higher own-price elasticities.

While both methods give confirming results, in the DM methodology the choice of distances is ambiguous and depends on prior judgments about the data, and trial and error. Consequently, it requires a cumbersome elimination method to test for different distance combinations in order to determine the best measure. Our approach is practical, eliminates the need to search for significant characteristics and has a stronger theoretical foundation. Thus we alleviate the ambiguity while significantly reducing the number of parameters. However, it is not necessarily the case that distances based on hedonic metrics give the best distance measurement. Future research is needed to find the optimal set of distances to model consumer's allocation.

While it is preferable to use individual level purchase data to analyze their purchase behavior using traditional demand models, it is impossible to use distance metrics at the household level. Since one of our primary motivations was to compare both methods, we used aggregate data. However, in the future, it will be possible to use hedonic metrics to segment consumers into different groups and create a separate hedonic space for each consumer group. Therefore, we can incorporate consumer demographics into the hedonic equations and estimate the elasticities based on not only product characteristics but also consumer characteristics in a simple manner. In this case, products targeting specific consumer profiles will be closely located in hedonic space which will result in higher cross-price elasticities between these products. That sort of research can further assist producers and marketers of differentiated products to effectively price and position their products within the market.

The metric model applied in this paper can be applied to any market where product differentiation exists. Although we applied our model to fluid milk products at retail level, our model makes it possible to estimate the elasticities between differentiated products in any market. In many industries such as the automotive industry or household appliances, we observe both close and distant competition which is difficult to model. Using hedonic metrics, we can accommodate this behavior in a robust and simple way.

**Table 1**
**Mean Attribute Values by Milk Type**

| Attributes | Milk Types | | | | | |
|---|---|---|---|---|---|---|
| | All | 0% | 1% | 2% | 3.25% | SOY |
| Organic Claim (%) | 2.44 | 0.58 | 1.01 | 0.48 | 0.61 | 63.81 |
| | (15.49) | (7.56) | (9.9) | (6.89) | (7.76) | (48.05) |
| Promotion (%) | 8.67 | 7.90 | 9.29 | 9.76 | 6.27 | 14.16 |
| | (28.09) | (26.96) | (29.03) | (29.67) | (24.23) | (34.86) |
| CFLF (%) | 5.04 | 3.87 | 1.19 | 2.21 | 0.94 | 98.99 |
| | (21.94) | (19.29) | (10.83) | (14.69) | 9.62 | (10.02) |
| Vitamin-Mineral Label (%) | 96.79 | 98.69 | 96.88 | 97.53 | 97.18 | 68.99 |
| | (17.48) | (11.34) | (17.39) | (15.53) | (16.56) | (46.25) |
| Protein per Serving | 8.40 | 8.81 | 8.41 | 8.54 | 8.04 | 5.07 |
| | (0.72) | (0.22) | (0.11) | (0.38) | (0.03) | (1.56) |
| Carbohydrate per Serving | 13.09 | 12.69 | 13.54 | 13.15 | 12.23 | 18.88 |
| | (3.40) | (1.54) | (3.67) | (3.71) | (3.56) | (4.01) |
| Fat per Serving | 3.68 | 0.53 | 2.26 | 4.77 | 8.15 | 2.41 |
| | (2.73) | (0.17) | (0.64) | (0.36) | (0.08) | (1.29) |
| Cholesterol DRI | 4.89 | 1.68 | 4.01 | 6.55 | 8.44 | 0.10 |
| | (2.68) | (0.09) | (0.49) | (0.74) | (0.41) | (0.98) |
| Sodium DRI | 4.97 | 5.29 | 4.75 | 5.24 | 4.30 | 4.40 |
| | (0.67) | (0.48) | (0.49) | (0.57) | (0.51) | (0.95) |
| Vitamin-Mineral DRI | 11.50 | 12.02 | 11.35 | 11.92 | 10.54 | 8.71 |
| | (1.07) | (0.16) | (0.11) | (1.25) | (0.06) | (1.63) |
| Servings per Package | 14.71 | 14.93 | 14.69 | 15.63 | 13.66 | 8.40 |
| | (4.72) | (10.25) | (9.02) | (10.48) | (9.68) | (5.73) |
| Unique UPC Codes | 860 | 142 | 200 | 225 | 262 | 31 |

Note: All DRI values are calculated as percentages of values provided per serving. Values in parenthesis represent standard deviations.

**Table 2**
**Price, Quantity and Market Share Statistics by Milk Type**

| Milk Type: | 2% | 0% | 3.25% | 1% | 0% |
|---|---|---|---|---|---|
| **Quantities** | | | | | |
| Mean | 12925.71 | 10524.45 | 6345.95 | 6526.62 | 615.31 |
| Std Dev | 725.88 | 623.34 | 611.65 | 409.16 | 142.79 |
| **Prices** | | | | | |
| Mean | 17.82 | 17.45 | 19.35 | 18.33 | 34.85 |
| Std Dev | 1.33 | 1.14 | 1.59 | 1.26 | 2.02 |
| **Market Shares** | | | | | |
| Mean | 34.00% | 27.13% | 18.07% | 17.66% | 3.14% |
| Std Dev | 0.96% | 0.88% | 1.02% | 0.76% | 0.58% |

Note: Prices are measured in cents per serving. A gallon of milk has 16 servings in a single package.

**Table 3**
**Rotterdam Model, Elasticity Estimates at Sample Means, N=208**

| | Hicksian (Compensated) Elasticities | | | | | | | | | |
|---|---|---|---|---|---|---|---|---|---|---|
| | PERCENT2 | | SKIM | | FULLFAT | | PERCENT1 | | SOY | |
| | Estimate | S.E. | Estimate | S.E. | Estimate | S.E. | Estimate | S.E. | Estimate | S.E. |
| PERCENT2 | -0.4781* | 0.150 | 0.2708* | 0.111 | 0.1196 | 0.096 | 0.1704* | 0.084 | -0.0052 | 0.052 |
| SKIM | 0.33953* | 0.139 | -0.575* | 0.166 | 0.0912 | 0.113 | 0.0877 | 0.086 | 0.0757 | 0.070 |
| FULLFAT | 0.22503 | 0.180 | 0.1369 | 0.169 | -0.5249* | 0.159 | 0.0741 | 0.137 | 0.1523 | 0.084 |
| PERCENT1 | 0.32803* | 0.162 | 0.1347 | 0.133 | 0.0759 | 0.140 | -0.6361* | 0.172 | -0.0825 | 0.106 |
| SOY | -0.0562 | 0.57 | 0.6558 | 0.608 | 0.8793 | 0.486 | -0.4653 | 0.599 | -1.0135* | 0.261 |

| | Marshallian (Uncompensated) Elasticities | | | | | | | | | |
|---|---|---|---|---|---|---|---|---|---|---|
| | PERCENT2 | | SKIM | | FULLFAT | | PERCENT1 | | SOY | |
| | Estimate | S.E. | Estimate | S.E. | Estimate | S.E. | Estimate | S.E. | Estimate | S.E. |
| PERCENT2 | -0.8179* | 0.148 | -0.0002 | 0.112 | -0.061 | 0.097 | -0.0061 | 0.085 | -0.0365 | 0.057 |
| SKIM | -0.047 | 0.140 | -0.8832* | 0.170 | -0.1149 | 0.111 | -0.1130 | 0.087 | 0.0401 | 0.075 |
| FULLFAT | -0.054 | 0.177 | -0.0857 | 0.172 | -0.6733* | 0.160 | -0.0708 | 0.137 | 0.1266 | 0.083 |
| PERCENT1 | 0.0061 | 0.160 | -0.1221 | 0.138 | -0.0952 | 0.141 | -0.8031* | 0.173 | -0.1121 | 0.101 |
| SOY | -0.4497 | 0.593 | 0.3419 | 0.611 | 0.6701 | 0.493 | -0.6697 | 0.602 | -1.049* | 0.254 |

| | Expenditure Elasticities | | | | | | | | | |
|---|---|---|---|---|---|---|---|---|---|---|
| | PERCENT2 | | SKIM | | FULLFAT | | PERCENT1 | | SOY | |
| | Estimate | S.E. | Estimate | S.E. | Estimate | S.E. | Estimate | S.E. | Estimate | S.E. |
| Estimate | 0.9993* | 0.068 | 1.1366* | 0.085 | 0.8210* | 0.0915 | 0.9467* | 0.0923 | 1.1571* | 0.353 |

Note: an asterisk (*) denotes significant at the 5% level; PERCENT2 denotes 2%, SKIM 0%, FULL FAT 3.25%, PERCENT1 1%, and SOY soymilk. Results are derived from (1).

**Table 4**
**Single Dimensional Distance Parameter Estimates**

| **One Dimensional** | | | |
|---|---|---|---|
| Distance | Estimate | S.E. | Log L |
| $\lambda_1$ : Market Share | 0.0239* | 0.0039 | 2729 |
| $\lambda_2$ : Fat Content | 0.015* | 0.0029 | 2726 |
| $\lambda_3$ : Organic Percentage | 0.0312* | 0.0045 | 2735 |
| $\lambda_4$ : Size | 0.0197* | 0.0032 | 2730 |
| **Two Dimensional** | | | |
| Distance | Estimate | S.E. | Log L |
| $\lambda_5$ : Fat-Organic | 0.0336* | 0.0055 | 2730 |
| $\lambda_6$ : Fat-Size | 0.0236* | 0.0037 | 2731 |
| $\lambda_7$ : Organic-Size | 0.0333* | 0.0048 | 2735 |
| **Three Dimensional** | | | |
| Distance | Estimate | S.E. | Log L |
| $\lambda_8$ : Fat-Organic-Size | 0.0356* | 0.0058 | 2731 |

Note: an asterisk (*) denotes significant at the 5% level; Log L is the log of the likelihood function. Results are derived from (2).

**Table 5**
**DM Distance Parameter Estimates and Elasticities for Rotterdam Model**

**Distance Parameter Estimates**
**Full Set of Distances**

|  | Estimate | S.E |  |  |
|---|---|---|---|---|
| $\lambda_1$ : Market Share | -0.0745 | 0.1188 | LogL | 2740.689 |
| $\lambda_2$ : Fat Content | -0.079 | 0.260 | Parameters |  |
| $\lambda_3$ : Organic Percentage | 1.187 | 1.343 | Estimated | 23 |
| $\lambda_4$ : Size | -4.232 | 5.756 | N | 208 |
| $\lambda_5$ : Fat-Organic | 7.107 | 15.305 | AIC | -5435.378 |
| $\lambda_6$ : Fat-Size | 5.998 | 7.8858 | BIC | -5428.063 |
| $\lambda_7$ : Organic-Size | -2.691 | 3.1506 |  |  |
| $\lambda_8$ : Fat-Organic-Size | -7.095 | 15.478 |  |  |
| $NN_{FO}$: NN in Fat-Organic Space | -0.014 | 0.036 |  |  |
| $NN_{FS}$: NN in Fat-Size Space | -0.034 | 0.055 |  |  |
| $NN_{FOS}$: NN in Fat-Organic-Size Space | -0.038 | 0.027 |  |  |
| $\beta_0$: Own-Price Coefficient | -0.242 | 0.457 |  |  |
| $\beta_1$: Market Share Own-Price Term | 0.1046 | 0.649 |  |  |
| $\beta_2$: Fat Content Own-Price Term | 0.001 | 0.012 |  |  |
| $\beta_3$: Organic Percentage Own-Price Term | 7.632 | 33.906 |  |  |

**Distance Version: F/O/NN$_{FO}$**

|  | Estimate | S.E |  |  |
|---|---|---|---|---|
| $\lambda_2$: Fat Content | -0.0026 | 0.0079 | LogL | 2734.468 |
| $\lambda_3$: Organic Percentage | 0.0418* | 0.0103 | Parameters |  |
| $NN_{FO}$: NN in Fat-Organic Space | -0.0194 | 0.0165 | Estimated | 15 |
| $\beta_0$: Own-Price Coefficient | -0.2838* | 0.143 | N | 208 |
| $\beta_1$: Market Share Own-Price Term | 0.3959 | 0.2963 | AIC | -5438.936 |
| $\beta_2$: Fat Content Own-Price Term | 0.0054 | 0.0051 | BIC | -5434.165 |
| $\beta_3$: Organic Percentage Own-Price Term | 8.8648 | 9.1188 |  |  |

Note: an asterisk (*) denotes significance at the 5% level. A product type is the Nearest Neighbor (NN) of another if it is the closest neighbor in the distance space.

## Table 5. (Cont.)
## DM Distance Parameter Estimates and Elasticities for Rotterdam Model

**Rotterdam Model (All Distances)**

**Marshallian (Uncompensated) Elasticities**

|          | PERCENT2 |      | SKIM     |      | FULLFAT  |      | PERCENT1 |      | SOY      |      |
|----------|----------|------|----------|------|----------|------|----------|------|----------|------|
|          | Estimate | S.E. | Estimate | S.E. | Estimate | S.E. | Estimate | S.E. | Estimate | S.E. |
| PERCENT2 | -0.6787✓* | 0.13 | -0.1602✓ | 0.11 | 0.0628✓ | 0.11 | -0.1348✓ | 0.12 | 0.0079✓ | 0.12 |
| SKIM     | -0.1122✓ | 0.14 | -0.8869✓* | 0.18 | -0.0525✓ | 0.10 | -0.0807✓ | 0.08 | 0.0416✓ | 0.07 |
| FULLFAT  | -0.1205✓ | 0.22 | -0.0076✓ | 0.16 | -0.6510✓* | 0.17 | -0.2439✓ | 0.34 | 0.3173X | 0.31 |
| PERCENT1 | 0.0745✓ | 0.18 | -0.0643✓ | 0.13 | -0.0595✓ | 0.16 | -0.8157✓ | 0.48 | 0.1003X | 0.11 |
| SOY      | -0.4181✓ | 0.61 | 0.3516✓ | 0.61 | 0.6836✓ | 0.49 | -0.6988✓ | 0.61 | -1.0743✓* | 0.27 |

**Expenditure Elasticities**

|          | PERCENT2 |      | SKIM     |      | FULLFAT  |      | PERCENT1 |      | SOY      |      |
|----------|----------|------|----------|------|----------|------|----------|------|----------|------|
|          | Estimate | S.E. | Estimate | S.E. | Estimate | S.E. | Estimate | S.E. | Estimate | S.E. |
| Estimate | 1.0119✓* | 0.07 | 1.123✓* | 0.09 | 0.8606✓* | 0.09 | 0.9029✓* | 0.09 | 1.1561✓* | 0.35 |

**Rotterdam Model (Distance Version: F/O/NN$_{FO}$)**

**Marshallian (Uncompensated) Elasticities**

|          | PERCENT2 |      | SKIM     |      | FULLFAT  |      | PERCENT1 |      | SOY      |      |
|----------|----------|------|----------|------|----------|------|----------|------|----------|------|
|          | Estimate | S.E. | Estimate | S.E. | Estimate | S.E. | Estimate | S.E. | Estimate | S.E. |
| PERCENT2 | -0.5763✓* | 0.09 | -0.1533✓* | 0.03 | -0.0631✓* | 0.03 | -0.1179✓* | 0.04 | 0.003✓ | 0.02 |
| SKIM     | -0.2347✓* | 0.04 | -0.7586✓* | 0.11 | -0.0532✓ | 0.04 | -0.1251✓* | 0.05 | 0.0115✓ | 0.01 |
| FULLFAT  | -0.1692✓* | 0.08 | -0.0007✓ | 0.05 | -0.7813✓* | 0.13 | 0.0749✓ | 0.05 | 0.044✓* | 0.02 |
| PERCENT1 | -0.0928✓ | 0.06 | -0.1387✓ | 0.07 | 0.0594✓ | 0.05 | -0.7955✓* | 0.12 | 0.0406✓ | 0.03 |
| SOY      | -0.6395✓ | 0.41 | 0.0908✓ | 0.15 | 0.1952✓ | 0.13 | 0.1889✓ | 0.16 | -0.996✓* | 0.25 |

**Expenditure Elasticities**

|          | PERCENT2 |      | SKIM     |      | FULLFAT  |      | PERCENT1 |      | SOY      |      |
|----------|----------|------|----------|------|----------|------|----------|------|----------|------|
|          | Estimate | S.E. | Estimate | S.E. | Estimate | S.E. | Estimate | S.E. | Estimate | S.E. |
| Estimate | 1.0041✓* | 0.07 | 1.1292✓* | 0.08 | 0.8369✓* | 0.09 | 0.9320✓* | 0.09 | 1.1605✓* | 0.35 |

Note: Asterisk (*) and double asterisk (**) denote variables significant at 5% level and 10% respectively; a check mark (✓) indicates that the estimated parameter is within the 95% confidence interval of the original model. Parameter estimates and elasticities are derived from (7). (F-O-NN$_{FO}$) denotes Fat, Organic, and NN for Fat-Organic space. A product type is the Nearest Neighbor (NN) of another if it is the closest neighbor in the distance space.

**Table 6**
**Hedonic Attribute Estimates**

| | Linear | | Semi-Log | |
|---|---|---|---|---|
| *Variable* | *Parameter Estimate* | *Std. Error* | *Parameter Estimate* | *Std. Error* |
| Intercept | -20.627 | 0.367 | 1.667 | 0.018 |
| Marketing Organic Claim | 10.962 | 0.090 | 0.428 | 0.004 |
| Marketing Soy Dummy | -9.351 | 0.155 | -0.367 | 0.008 |
| Marketing Promotion Dummy | -1.583 | 0.026 | -0.100 | 0.001 |
| Marketing Lactose Cholesterol Free | 23.537 | 0.069 | 0.857 | 0.003 |
| Marketing Vitamin Mineral Index | 4.497 | 0.077 | 0.140 | 0.004 |
| Nutrient Protein Content (g) | 2.734 | 0.042 | 0.085 | 0.002 |
| Nutrient Carb Content (g) | 0.991 | 0.007 | 0.033 | 0.000 |
| Nutrient Lipid (Fat) Content (g) | 0.861 | 0.014 | 0.032 | 0.001 |
| Nutrient Cholesterol DRI Max | -0.358 | 0.014 | -0.011 | 0.001 |
| Nutrient Sodium DRI Max | -2.010 | 0.027 | -0.060 | 0.001 |
| Nutrient Vit-Min Percentage Index | 0.789 | 0.016 | 0.020 | 0.001 |
| Purchased Serving Size Quantity | -0.106 | 0.001 | -0.005 | 0.001 |
| Adjusted R-Square Value | 0.3666 | | 0.2872 | |

Note: All estimates are significant at 1% level. Linear parameter estimates are derived from (8) whereas semi-log parameter estimates are derived from (10).

**Table 7**
**Hedonic Metric Parameter Estimates and Elasticities**

| Rotterdam Model (Semi-Log) | | | Rotterdam Model (Linear) | | |
|---|---|---|---|---|---|
| Parameter | Estimate | S.E | Parameter | Estimate | S.E |
| $\lambda_h$* | 0.0453 | 0.009 | $\lambda_h$* | 0.0488 | 0.011 |
| $\lambda_{nn}$** | -0.0281 | 0.014 | $\lambda_{nn}$** | -0.0290 | 0.014 |
| $\beta_0$ | 0.2147 | 0.836 | $\beta_0$ | -0.1237 | 0.336 |
| $\beta_1$ | 0.1218 | 0.176 | $\beta_1$ | 0.1679 | 0.169 |
| $\beta_2$ | -1.0149 | 2.443 | $\beta_2$ | -0.0305 | 0.953 |
| LogL | 2733.933 | | Log L | 2733.662 | |
| Parameters Estimated | 13 | | Parameters Estimated | 13 | |
| N | 208 | | N | 208 | |
| AIC | -5441.866 | | AIC | -5441.324 | |
| BIC | -5437.7312 | | BIC | -5437.1892 | |

**Hedonic Metric Approximated Elasticities for Rotterdam Model**
**Semi-Log**

**Marshallian (Uncompensated) Elasticities**

| | PERCENT2 | | | SKIM | | | FULLFAT | | | PERCENT1 | | | SOY | | |
|---|---|---|---|---|---|---|---|---|---|---|---|---|---|---|---|
| | Estimate | | S.E. | Estimate | | S.E. | Estimate | | S.E. | Estimate | | S.E. | Estimate | | S.E. |
| PERCENT2 | -0.5822 | ✓* | 0.073 | -0.1543 | ✓* | 0.032 | -0.0622 | ✓* | 0.029 | -0.1396 | ✓* | 0.026 | 0.0146 | ✓ | 0.01 |
| SKIM | -0.2376 | ✓* | 0.041 | -0.6617 | ✓* | 0.069 | -0.0685 | ✓* | 0.031 | -0.1516 | ✓* | 0.033 | 0.0211 | ✓ | 0.012 |
| FULLFAT | -0.2159 | ✓* | 0.055 | -0.0227 | ✓ | 0.045 | -0.7638 | ✓* | 0.102 | 0.0674 | ✓ | 0.045 | 0.0573 | ✓* | 0.017 |
| PERCENT1 | -0.0850 | ✓ | 0.057 | -0.1778 | ✓* | 0.052 | 0.0522 | ✓ | 0.049 | -0.7481 | ✓* | 0.099 | 0.0575 | ✓* | 0.018 |
| SOY | -0.8041 | ✓* | 0.397 | 0.1659 | ✓ | 0.138 | 0.2660 | ✓* | 0.118 | 0.2771 | ✓* | 0.118 | -1.1004 | ✓* | 0.223 |

**Expenditure Elasticities**

| | PERCENT2 | | | SKIM | | | FULLFAT | | | PERCENT1 | | | SOY | | |
|---|---|---|---|---|---|---|---|---|---|---|---|---|---|---|---|
| | Estimate | | S.E. | Estimate | | S.E. | Estimate | | S.E. | Estimate | | S.E. | Estimate | | S.E. |
| | 1.0016 | ✓* | 0.066 | 1.1314 | ✓* | 0.084 | 0.8358 | ✓* | 0.089 | 0.9284 | ✓* | 0.09 | 1.1958 | ✓* | 0.349 |

**Linear**

**Marshallian (Uncompensated) Elasticities**

| | PERCENT2 | | | SKIM | | | FULLFAT | | | PERCENT1 | | | SOY | | |
|---|---|---|---|---|---|---|---|---|---|---|---|---|---|---|---|
| | Estimate | | S.E. | Estimate | | S.E. | Estimate | | S.E. | Estimate | | S.E. | Estimate | | S.E. |
| PERCENT2 | -0.5627 | ✓* | 0.071 | -0.151 | ✓* | 0.035 | -0.0673 | ✓* | 0.03 | -0.1377 | ✓* | 0.022 | 0.0172 | ✓ | 0.011 |
| SKIM | -0.2416 | ✓* | 0.039 | -0.6368 | ✓* | 0.068 | -0.0846 | ✓* | 0.028 | -0.1592 | ✓* | 0.033 | 0.0254 | ✓ | 0.013 |
| FULLFAT | -0.2344 | ✓* | 0.054 | -0.0431 | ✓ | 0.043 | -0.7254 | ✓* | 0.089 | 0.0560 | ✓ | 0.044 | 0.0639 | ✓* | 0.02 |
| PERCENT1 | -0.2448 | ✓* | 0.046 | -0.0222 | ✓ | 0.06 | 0.0401 | ✓ | 0.051 | -0.755 | ✓* | 0.099 | 0.0687 | ✓* | 0.022 |
| SOY | 0.1139 | ✓ | 0.168 | 0.2035 | ✓ | 0.153 | 0.3023 | ✓* | 0.133 | -0.5883 | ✓ | 0.349 | -1.2373 | ✓* | 0.21 |

**Expenditure Elasticities**

| | PERCENT2 | | | SKIM | | | FULLFAT | | | PERCENT1 | | | SOY | | |
|---|---|---|---|---|---|---|---|---|---|---|---|---|---|---|---|
| | Estimate | | S.E. | Estimate | | S.E. | Estimate | | S.E. | Estimate | | S.E. | Estimate | | S.E. |
| | 0.9902 | ✓* | 0.068 | 1.1439 | ✓* | 0.086 | 0.8347 | ✓* | 0.09 | 0.9302 | ✓* | 0.092 | 1.2058 | ✓* | 0.346 |

Note: an asterisk (*) denotes significant at the 5% level; a check mark (✓) indicates that the estimated parameter is within the 95% confidence interval of the original model. Parameter estimates and elasticities are derived from (14).